\documentclass[11pt,preprint]{aastex}
\newcommand{\be}{\begin{equation}}
\newcommand{\ba}{\begin{eqnarray}}
\newcommand{\ee}{\end{equation}}
\newcommand{\ea}{\end{eqnarray}}
\newcommand{\etal}{et al.\ }
\newcommand{\etalb}{et al.}
\newcommand{\Omm}{\Omega_m}

\begin{document}
\title{The Star Formation Rate Intensity Distribution Function --- 
Comparison of Observations with Hierarchical Galaxy Formation}

\author{Rennan Barkana}
\affil{School of Physics and Astronomy, Tel Aviv University, Tel Aviv 69978,
ISRAEL}
\email{barkana@wise.tau.ac.il}

\begin{abstract}
Recently, \citet{l02} have measured the distribution of star formation
rate intensity in galaxies at various redshifts. This data set has a
number of advantages relative to galaxy luminosity functions; the
effect of surface-brightness dimming on the selection function is
simpler to understand, and this data set also probes the size
distribution of galactic disks. We predict this function using
semi-analytic models of hierarchical galaxy formation in a
$\Lambda$CDM cosmology. We show that the basic trends found in the
data follow naturally from the redshift evolution of dark matter
halos. The data are consistent with a constant efficiency of turning
gas into stars in galaxies, with a best-fit value of $2\%$, where dust
obscuration is neglected; equivalently, the data are consistent with a
cosmic star formation rate which is constant to within a factor of two
at all redshifts above two. However, the practical ability to use this
kind of distribution to measure the total cosmic star formation rate
is limited by the predicted shape of an approximate power law with a
smoothly varying power, without a sharp break.
\end{abstract}

Key Words: galaxies: high-redshift, cosmology: theory, 
galaxies: formation

\vspace{-.1in}
PACS: 98.80.-k, 98.62.Ai

\section{Introduction}
\label{intro}

One of the major goals of the study of galaxy formation is to achieve
an observational determination and a theoretical understanding of the
cosmic star formation history. Previous measurements have generally
found a factor of $\sim 4-10$ increase in the cosmic star formation
rate (henceforth SFR) from redshift $z=0$ out to $z \sim 1$--2, with
the cosmic SFR remaining roughly constant at higher redshifts out to
$z \sim 4$, once approximate corrections are included for
incompleteness or for the effect of dust extinction.

In general, estimates of the SFR apply locally-calibrated correlations
between emission in particular lines or wavebands and the total
SFR. The observational picture is based on a large number of
observations in different wavebands. These include various
ultraviolet/optical/near-infrared observations \citep[e.g.,]
[]{m96,l96,mpd98,t98,p98,csb99,g99,f99,s99,hcs00}. At the shortest
wavelengths, the extinction correction may be large (a factor of $\sim
5$ at the highest redshifts) and is still highly uncertain. At longer
wavelengths, the star formation history has been reconstructed from
submillimeter observations \citep{b99,h98} and radio observations
\citep{c98}; in this range of the spectrum, large uncertainties remain
because of the insufficiency of current observational constraints on
the spectral shape of the galaxies' dust emission.

Hierarchical models have been used in many papers to match
observations on star formation at $z \la 4$ \citep[e.g.,][]
{b98,kc98,spf01}. This comparison should become increasingly
profitable for our physical understanding of galaxy formation as
observations probe the galaxy population more completely at low
redshift and also push towards high redshift.

Recently, \citet{l02} introduced a new data set, which takes advantage
of the deepest images taken by the Hubble Space Telescope (HST),
namely the Hubble Deep Field (HDF), and the Hubble Deep Field South
(HDF-S) WFPC2 and NICMOS fields. These fields contain a fairly large
number of galaxies, including a good fraction at high
redshift. Combining a large variety of space- and ground-based optical
and infrared images of these fields, \citet{l02} found photometric
redshifts for $\sim 3000$ galaxies. They measured redshifts using a
sequence of six spectral templates which account for different galaxy
types. Their redshift technique has been checked with spectroscopic
measurements at $z<6$, yielding a relative root-mean-square dispersion
of 0.065 in $1+z$; however, the check applies mainly to $z=0$--1.2 and
$z=2.2$-3.5, with only a small number of checks outside these redshift
ranges \citep{y00}. Another limitation of the current data is that the
fields are relatively small, and are not randomly selected.  The HDF
was chosen to be in a relatively empty field, while the HDF-S was
chosen to be near a $z=2.2$ quasar. Thus, cosmic variance may be
significant, although it should be greatly suppressed by the large
redshift bins that are used in the analysis.

The data analysis of \citet{l02} presents several novelties compared
to optical and infrared determinations of the galaxy luminosity
function. Since most galaxies are well-resolved in the HST images,
\citet{l02} divide each galaxy into individual pixels, and measure
the SFR intensity $x$ in each pixel, in units of $M_{\odot}$ yr$^{-1}$
kpc$^{-2}$, where all quantities are proper. They then add up the
proper areas of all pixels, within a given redshift range, with SFR
intensity in the interval $x$ to $x+dx$. Dividing by the comoving
volume in the redshift bin, and by $dx$, they obtain $h(x)$, the SFR
intensity distribution function, expressed in units of proper kpc$^2$
per comoving Mpc$^3$ per unit of $x$, i.e., kpc$^2$ Mpc$^{-3}/
(M_{\odot}$ yr$^{-1}$ kpc$^{-2})$. Once $h(x)$ has been obtained, the
total cosmic SFR per comoving volume is simply $\int x\, h(x)\, dx$.

Measuring the SFR in pixels offers a number of advantages compared to
measuring the total SFR in a galaxy. First, as noted by \citet{l02},
the total luminosity of a galaxy is in practice not very well defined,
because the luminosity is integrated out to a radius where the surface
brightness drops below the noise, and this radius depends strongly on
redshift due to cosmological surface brightness dimming. In addition,
the effect of surface brightness limits on the selection function is
non-trivial in the case of galaxy luminosity functions; specifically,
whether or not a galaxy is detected does not depend only on its
luminosity, but also on its size and its orientation relative to the
line of sight. On the other hand, since each pixel has a definite,
known angular size, the selection function is simple, i.e., there is a
minimum $x$ that can be detected within the pixel, as a function of
redshift (and of position on the CCD). For this data set, the size
distribution of galaxies does not enter as a nuisance in the data
analysis, instead it enters as an important element of any theoretical
model, an element which directly affects the predicted function
$h(x)$.

\citet{l02} obtained the function $h(x)$ over a limited range of $x$
values in each of ten redshift bins. They then fit the curves to a
broken power law model, allowing one of the parameters of the model to
vary with redshift. The choice of a broken power law and of the
allowed types of variation with redshift was motivated by the
appearance of the data, not by any physical model of galaxy
formation. Due to surface brightness dimming, at high redshift the
observations can only detect the upper end of pixels, i.e., those with
the highest SFR intensities. Indeed, at all redshifts $z>2$, most of
the cosmic SFR occurs at $x$ values that are below the detection
limit, and thus the total cosmic SFR is sensitive to an extrapolation
which depends on the assumed shape of the function $h(x)$. In this
paper, we confront the data with a semi-analytic model based on the
theoretical understanding of hierarchical galaxy formation in a
$\Lambda$CDM cosmology. We examine whether the model can explain the
overall trends in the data, and we use the predicted,
physically-motivated shape of $h(x)$ to carry out a measurement of the
cosmic SFR.

The basic theoretical framework in which the matter content of the
universe is dominated by CDM has recently received a major
confirmation from measurements of the cosmic microwave background
\citep{Boomerang,Maxima,Dasy}. Based primarily on these measurements, 
in this paper we assume a $\Lambda$CDM cosmology with parameters
$\Omm$ = 0.3, $\Omega_\Lambda$ = 0.7, $\Omega_b = 0.05$, $\sigma_8 =
0.8$, $n=1$, and $h=0.7$, where $\Omm$, $\Omega_\Lambda$, and
$\Omega_b$ are the total matter, vacuum, and baryon densities in units
of the critical density, $\sigma_8$ is the root-mean-square amplitude
of mass fluctuations in spheres of radius $8\ h^{-1}$ Mpc, and $n=1$
corresponds to a primordial scale-invariant power spectrum.

Throughout this paper we express results in physical units in
$\Lambda$CDM. Note that \citet{l02} calculated results in an $\Omm=1$
cosmology, and expressed most quantities in units of $h=1$. We convert
their measurements into our units and cosmology, using the redshift
distribution of their galaxy sample. Note that $x$ depends on redshift
but not on the cosmological matter content, since it is derived from
observations of surface brightness. However, $h(x)$ and the cosmic SFR
do change, through the change in both the proper area and the comoving
volume corresponding to a given solid angle. Using $h=0.7$ reduces
$h(x)$ by a factor of 1.4, and the conversion to $\Lambda$CDM causes
an additional decline by a factor which equals 1.3 for the $z=0$--0.5
bin, and rises up to 1.8 for all $z>2$ bins.

In the following section we present the details of our theoretical
model. Readers primarily interested in the comparison to the data may
go directly to the summary of the basic setup in \S3, the results in
\S 4 and the conclusions in \S 5.

\section{Theoretical Model}

In this section we describe a model which allows us to predict the SFR
intensity distribution function $h(x)$ in the context of hierarchical
galaxy formation in $\Lambda$CDM. Briefly, we require the distribution
of halo masses for halos which host galaxies, the SFR of each galaxy,
and the size distribution of galactic disks.

We assume that the abundance of halos is given by the Press-Schechter
model (\citealt{PS74}; recent corrections to this model are
insignificantly small for the current work). Galaxies form in halos in
which gas can accumulate and cool. At high redshift, gas can cool
efficiently in halos down to a virial temperature of $\sim 10^4$ K or
a circular velocity of $V_c\sim 16.5\ {\rm km\ s}^{-1}$ with atomic
cooling, which we assume to be the dominant cooling mechanism. Before
reionization, the IGM is cold and neutral, and these cooling
requirements set the minimum mass for halos which can host
galaxies. During reionization, however, when a volume of the IGM is
ionized by stars, the gas is heated to a temperature $T_{\rm IGM}\sim
1$--$2 \times 10^4$ K. We adopt a standard temperature of $T_{\rm
IGM}=1.5 \times 10^4$ K, and then the linear Jeans mass corresponds to
a virialized halo with a circular velocity of \be V_J=82
\left(\frac{T_{\rm IGM}} {1.5\times 10^4 {\rm K}}\right)^{1/2}\ {\rm
km\ s}^{-1}\ , \ee where this value is essentially independent of
redshift. Even halos well below the Jeans mass can pull in some gas
once the dark matter collapses to the virial overdensity. For
simplicity, we adopt a sharp cutoff associated with this suppression,
at a circular velocity of $V_c = V_J/2$, based on the results of
numerical simulations
\citep{tw96,qke96,whk97,ns97,ki00}. However, this pressure suppression
is not expected to cause an immediate suppression of the cosmic star
formation rate, since even after fresh gas infall is halted the gas
already in galaxies continues to produce stars, and mergers among
already-formed gas disks also trigger star formation. Indeed, after
reionization we assume that the minimum $V_c$ for hosting a galaxy
rises with time only gradually, so that the total gas fraction in
galaxies does not decline with time, as it physically must not [Some
gas already in galaxies before reionization does photoevaporate at
reionization, but this happens mostly in small halos which could not
have cooled via atomic cooling \citep{me99}]. We assume that
reionization occurs gradually, with \ion{H}{2} regions fully engulfing
the low density intergalactic medium at $z=6.5$, as suggested but not
implied by recent observations \citep{z6.3,z5.8,me02}. Note that while
we include aspects of reionization in the model, this mostly affects
the SFR at redshifts $z > 4.5$, where the photometric redshifts may be
less reliable; in this paper we focus on redshifts up to $z \sim 4$.

Once gas collects inside a halo and cools, it can collapse to high
densities and form stars (and perhaps also a mini-quasar). The ability
of stars to form is determined by gas accretion which, in a
hierarchical model of structure formation, is driven by mergers of
dark matter halos. Therefore, in order to determine the lifetime of a
typical source, we first define the age of gas in a given halo using
the average rate of mergers which built up the halo. Based on the
extended Press-Schechter formalism
\citep{lc93}, for a halo of mass $M$ at redshift $z$, the fraction of
the halo mass which by some higher redshift $z_2$ had already
accumulated in halos with galaxies is \be F_M(z,z_2) = {\rm Erfc}
\left(\frac{1.69/D(z_2)- 1.69/D(z)}{\sqrt{2 (S(M_{\rm
min}(z_2))-S(M))}} \right)\ , \ee where $D(z)$ is the linear growth
factor at redshift $z$, $S(M)$ is the variance on mass scale $M$
(defined using the linearly-extrapolated power spectrum at $z=0$), and
$M_{\rm min}(z_2)$ is the minimum halo mass for hosting a galaxy at
$z_2$ (as determined by the cutoff $V_c$ as was discussed above).

We estimate the total age of gas in the halo as the time since
redshift $z_2$ where $F_M(z,z_2) = 0.2$, so that most ($80\%$) of the
gas in the halo has fallen into galaxies only since then. Low-mass
halos form out of gas which has recently cooled for the first time,
while high-mass halos form out of gas which has already spent previous
time inside small galaxies. We emphasize that the age we have defined
here is not the formation age of the halo itself, but rather it is an
estimate for the total period during which the gas which is currently
in the halo has participated in star formation. However, the rate of
gas infall is not constant, and even within the galaxy itself, the gas
likely does not form stars at a uniform rate. Indeed, galaxies could
go through repeated cycles of a star formation burst followed by
feedback squelching, followed by another cycle of cooling,
fragmentation and star formation. The details involve complex
astrophysics, which are not understood even at low redshift, so we
account for the general possibility of bursting sources by adding a
parameter $\zeta \le 1$, the duty cycle. When $\zeta < 1$, compared to
$\zeta=1$ there are fewer sources (by a factor of $\zeta$) but each
has a larger SFR (by a factor of $1/\zeta$). Thus, $\zeta$ is a free
parameter which does not change the total cosmic SFR. In addition, the
cosmic SFR is proportional to the efficiency parameter $\eta$, which
is the fraction of gas in each galaxy which turns into stars. In
summary, the SFR of a galaxy in a halo of mass $M > M_{\rm min}(z)$ is
\be {\rm SFR} = \frac{\Omega_b}{\Omega_m}\, \frac{\eta}{\zeta}\,
\frac{M}{t_{\rm gas}}\ , \ee where $t_{\rm gas}$ is the gas
age as defined above.

A crucial element of the function $h(x)$ is its dependence on the size
distribution of galactic disks. The formation of galactic disks within
dark matter halos has been previously explored and compared to
observed properties of disks at low redshift \citep{fe80, dss97,
mmw98}. The observed distribution of disk sizes suggests that the
specific angular momentum of the disk is similar to that of the
halo. If they are assumed to indeed be equal, then for the simple
model of an exponential disk in a singular isothermal sphere halo, the
exponential scale radius of the disk is given by a fraction
$\lambda/\sqrt{2}$ of the halo virial radius, where $\lambda$ is the
dimensionless spin parameter of the halo
\citep{mmw98}.  The spin parameter distribution is approximately
independent of halo mass, environment, and cosmological parameters,
and approximately follows a lognormal distribution, \be p(\lambda)
d\lambda= \frac{1} {\sigma_{\lambda} \sqrt{2 \pi}} \exp \left [-\frac{
\ln^2(\lambda/ \bar{\lambda})}{2\sigma_{\lambda}^2} \right] 
\frac{d\lambda}{\lambda}\ , \label{eq:spin} \ee with $\bar{\lambda}=0.05$ 
and $\sigma_{\lambda}=0.5$ following \citet{mmw98}, who determined
these values based on the N-body simulations of \citet{w92}.

As a final element, we include a random orientation, i.e., a uniform
distribution of the angle of each galactic disk relative to the line
of sight to the observer. The distributions of both disk size and
orientation must be integrated over in order to determine the 
observable function $h(x)$. 

\section{Basic Setup}

Before discussing the results, we first summarize the essential
properties of the data and the model, focusing on their primary
advantages and disadvantages, some of which were discussed in the
previous sections.

The advantages of the data set of \citet{l02} are:
\begin{enumerate}
\item A large number of galaxies ($\sim$3000), all imaged at high
resolution with HST, and all with measured photometric redshifts.
\item Since most galaxies are well-resolved, the SFR can be measured 
within individual pixels rather than in entire galaxies. As a result,
the effect of surface-brightness dimming on the selection function is
simpler to understand than in the usual case of measuring the galaxy
luminosity function.
\item The measured distribution, which focuses on the SFR
intensity, is sensitive to the size distribution of galactic disks of
a given luminosity, and thus the data allows a test of this additional
aspect of theoretical models, unlike the galaxy luminosity function.
\end{enumerate}

The disadvantages of the data set are:
\begin{enumerate}
\item The observed fields are very deep but relatively small in 
angular size, and they were not randomly selected. Thus, cosmic
variance could be significant despite the large redshift bins that are
used.
\item Since galaxies are pixelized, this introduces some smoothing
relative to a comparison to a theoretical model, especially at high
redshift where the number of pixels per galaxy is relatively
small. Also, different photometric bands have different resolutions,
so the spectrum of each galaxy is assumed to be fixed; if the actual
spectrum varies among pixels in a given galaxy, then this affects the
inferred SFR.
\item The photometric redshift measurements use six galactic 
spectral templates, while these templates may in reality evolve with
redshift in unknown ways. However, this redshift technique has been
checked with spectroscopic measurements, though mainly at $z=0$--1.2
and $z=2.2$-3.5 \citep{y00}.
\end{enumerate}

The advantages of the theoretical model that we use are:
\begin{enumerate}
\item The model is based on the theoretical understanding
of hierarchical galaxy formation within a $\Lambda$CDM cosmology,
a cosmology which is strongly supported by recent observations.
\item The model assumes that galaxy formation is, on the whole,
driven by the hierarchical merging of dark matter halos. Specifically,
the timescale for star formation in a galaxy of a given mass and 
redshift is set by the infall history of the gas into halos.
\item Given the uncertainty, especially at high redshift, of the
detailed properties of star formation and feedback, the model includes
a reasonable flexibility with two free parameters, the efficiency of
star formation, and a duty cycle ($\le 1$) which allows for the
possibility of episodic bursts of star formation.
\item The model includes the distribution of galaxy disk sizes 
which is expected theoretically, given the distribution of the 
angular momenta of dark matter halos. It also includes a random
orientation of the disk relative to the line of sight.
\item The model accounts, in a simplified way, for the effect of 
reionization at redshift 6.5. This, however, mostly affects the star
formation rate at redshifts $z > 4.5$, where the photometric redshifts
may be less reliable.
\end{enumerate}

The disadvantages of the theoretical model are:
\begin{enumerate}
\item We do not include a distribution of possible merger histories
for a halo of a given mass and redshift. 
\item Our model focuses on disk galaxies, which should dominate star
formation, especially at high redshift where most of the gas in a
forming galaxy is falling in and forming stars for the first
time. However, ellipticals and strongly irregular or disrupted disks
are not correctly modeled in detail. Also, patchy star formation and
extinction may make disks appear irregular in the rest-frame
ultraviolet. Note, though, that the basic arguments for the star
formation timescales and for the characteristic sizes of galaxies are
very general, and should describe all galaxies sufficiently accurately
for an initial comparison with observational data.
\item We do not model feedback processes in detail, only through
the two free parameters. Also, we assume that the star formation
efficiency and the duty cycle are constant, while more complicated
models with additional free parameters are possible. E.g., we do not
model cooling in detail, but increasing virial temperatures and
decreasing densities should decrease the cooling efficiency at the
lowest redshifts. Our approach is to first compare the data to a model
which contains most of the relevant physics, in order to judge whether
this type of data could, in the future, constrain even more detailed
models.
\end{enumerate}

\section{Results}

To begin the comparison, we first note that the points of \citet{l02}
have a typical vertical $1-\sigma$ error of about a factor of 2 (up or
down). It is difficult to estimate the additional systematic errors
which arise from the just-noted limitations of the model and the data,
but they are likely at least as large as the assigned error bars. In
order to gauge the ability of the model to fit the data, we assume a
minimum $1-\sigma$ error of a factor of 2 on each $h(x)$ bin, but we
adopt the error bar of \citet{l02} on points where the error bar is
larger than this.

\begin{figure} 
\plotone{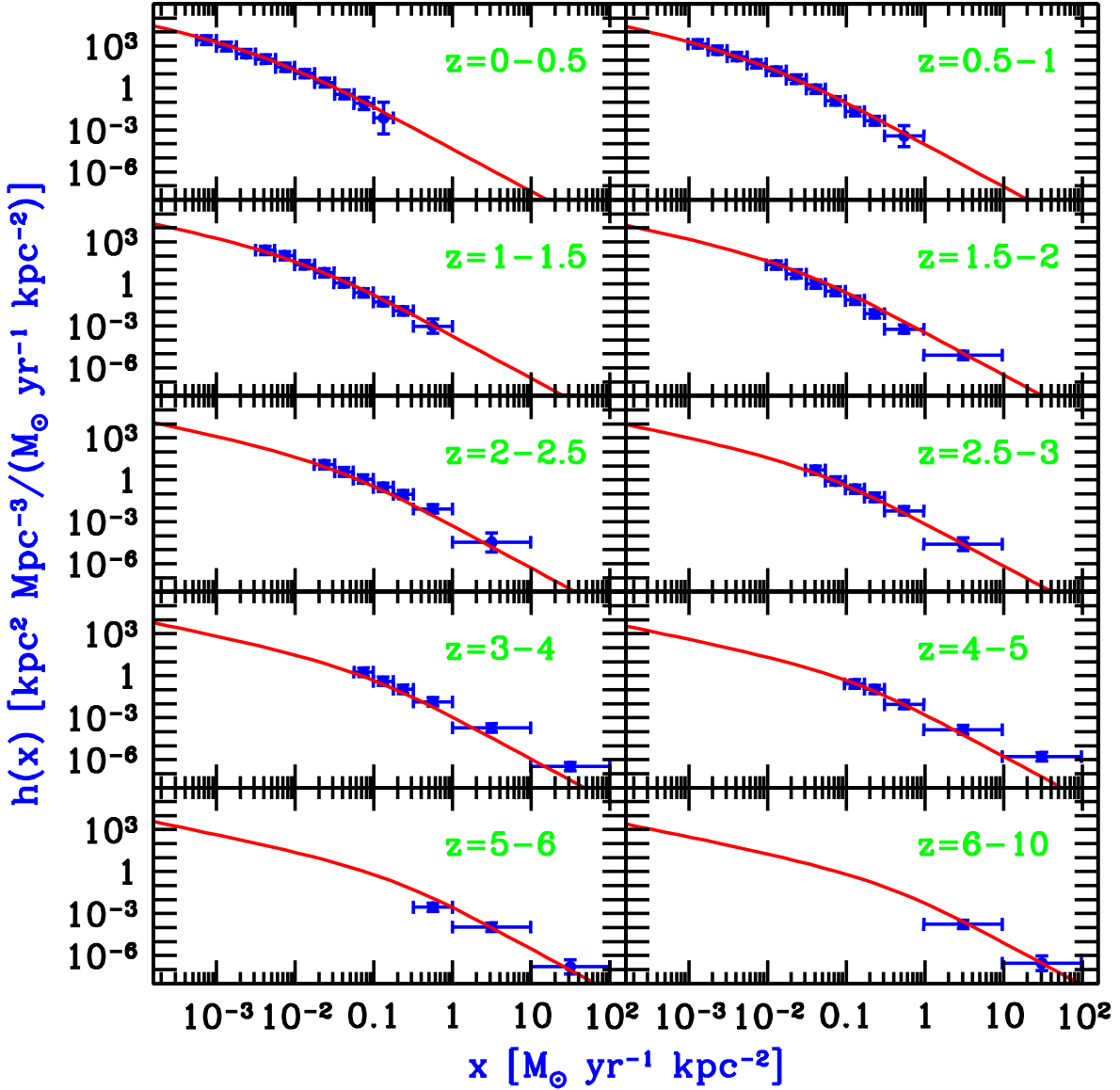} 
\caption{Comparison between our theoretical model and the data points 
of \citet{l02}. Shown is the SFR intensity distribution function
$h(x)$ versus SFR intensity $x$, for various redshift bins. Points
with error bars are taken from \citet{l02}, except that they are
converted into our units and cosmology, and the error bars are
slightly larger (see the text). Note that the horizontal brackets
indicate bins, not error bars. Solid curves show the best-fit
theoretical model (eq.\ \ref{best}).}
\label{fig-hofx}
\end{figure}

Figure \ref{fig-hofx} compares the best-fit theoretical model to the
data points of \citet{l02}, converted into our units and cosmology
(see the end of \S \ref{intro}). Note that while the figure shows the
smooth theoretical curves, in the fitting the theoretical predictions
were binned into the same $x$-bins as the data. The best-fit parameter
values and (formal) $1-\sigma$ errors are
\be \eta = 2.3^{+0.4}_{-0.3} \%\ , \ \ \ \ \zeta=17^{+7}_{-5} \%\ , 
\label{best}\ee
with $2-\sigma$ ranges of $1.8\% < \eta < 3.1\%$ and $9 < \zeta < 35
\%$.  The fit is acceptable, with a $\chi^2 = 59$ for 65 degrees of
freedom (67 data points and 2 parameters). The fit is better over the
redshift range with the most reliable data; fitting only the $z<4$
bins yields a $\chi^2 = 36$ for 55 degrees of freedom (with best-fit
parameters $\eta=2.4\%$ and $\zeta=20\%$). Thus, the data are
consistent with the theoretical model under the assumption of
parameters $\eta$ and $\zeta$ which do not evolve with redshift. Note
that a parameter $\zeta \sim 17\%$ is consistent with our assumption
of regular disks, if the star formation bursts are triggered mostly by
minor mergers which the disk may survive \citep[e.g.,][]{w96}.

Hierarchical galaxy formation provides a natural explanation for the
observed redshift evolution in the shape of $h(x)$. At high redshift,
there is a shift towards higher values of SFR intensity, with $h(x)$
increasing at large $x$ and decreasing at small $x$. Galaxies at high
redshift have lower typical masses than at low redshift, but the
typical value of $x$ increases nonetheless because of two additional
effects. First, high redshift disks are compact due to the increased
mean density of the universe and the proportional increase in the
typical density of virialized halos and of the disk galaxies forming
within them. The second effect results from the vigorous star
formation rates that are expected in high redshift galaxies despite
their small masses. The increased rates of star formation are caused
by the increased merger rate of dark matter halos. Since lower-mass
halos are forming at high redshift, the power spectrum at the relevant
mass scales is steeper, i.e., closer to the slope of $-3$ for which
all mass scales would collapse together. Thus, at high redshift, when
one mass scale collapses, a higher mass scale soon collapses as well,
and this corresponds to a vigorous merger rate [For a further
discussion of the redshift evolution of the size and surface
brightness of galactic disks, see \citet{me00}].

\begin{figure} 
\plotone{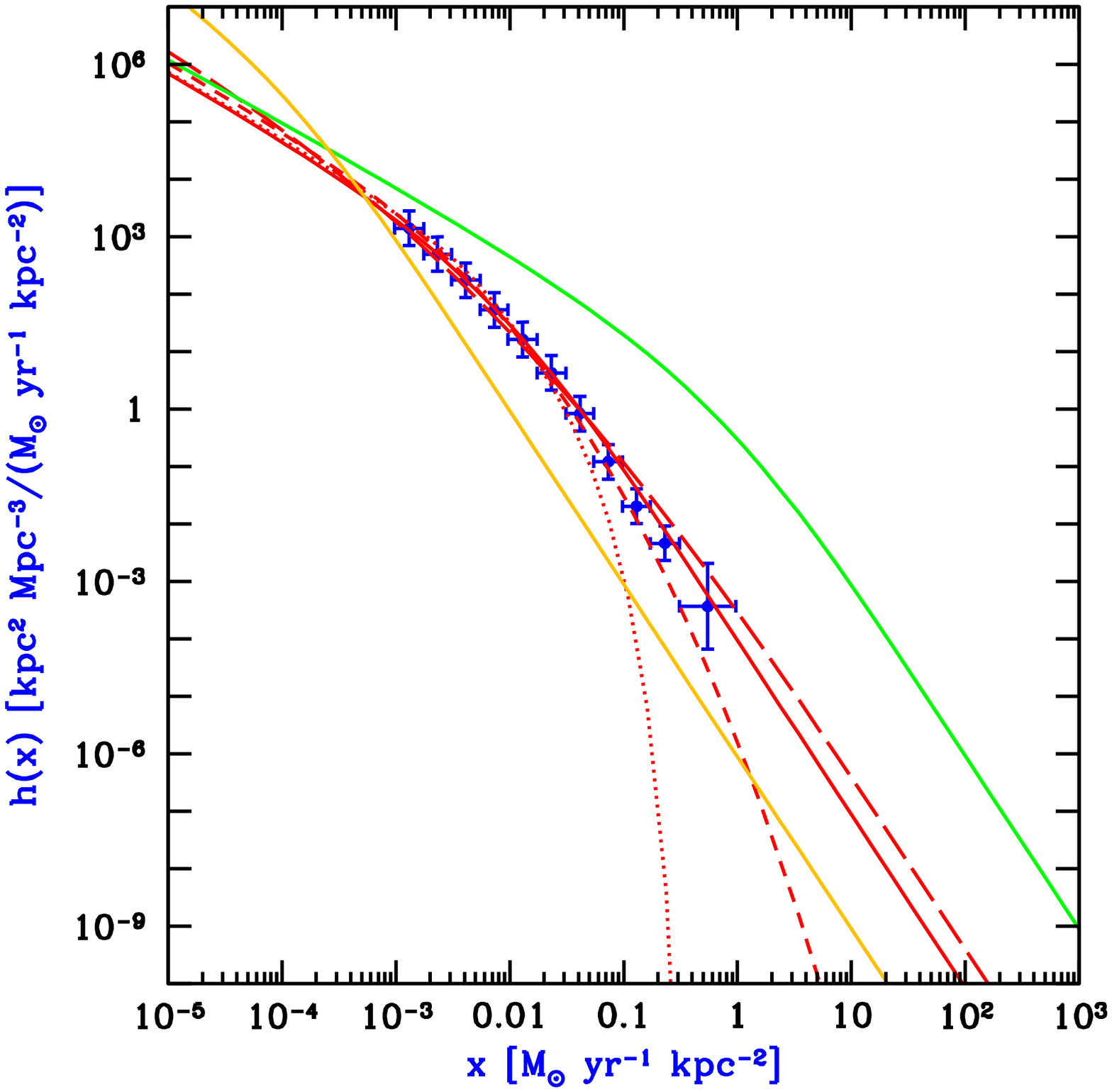} 
\caption{Theoretical models and data points in the $z=0.5-1$ redshift
bin. Shown is the SFR intensity distribution function $h(x)$ versus
SFR intensity $x$. Points with error bars are taken from \citet{l02}
except for modifications (see the caption of Figure \ref{fig-hofx}).
The solid curve which goes through the data points shows the
theoretical model with the parameters as in eq.\ (\ref{best}). The
other solid curves illustrate the effect of varying the parameters,
i.e., multiplying $\eta$ by a factor of 100 (highest curve at
$x=0.01$) or multiplying $\zeta$ by a factor of 100 (lowest curve at
$x=0.01$).  The three additional curves illustrate the effect on the
shape of $h(x)$ of increasing $\sigma_{\lambda}$ in eq.\
(\ref{eq:spin}) to unity (long-dashed curve), of assuming that all
galactic disks are face-on (short-dashed curve), or of assuming that
all galactic disks are face-on and have a halo spin parameter
$\lambda=0.05$ (dotted curve).}
\label{fig-z0p5}
\end{figure}

Figure \ref{fig-z0p5} illustrates how the shape of $h(x)$ is
determined in the theoretical model. If all galactic disks are assumed
to be face-on, and to also have a halo spin parameter equal to the
average value ($\lambda=0.05$), then $h(x)$ is driven by the shape of
the halo mass function and shows a steep drop-off at large values of
$x$. If the spin-parameter distribution is included, $h(x)$ is
smoother; the numerous galaxies in relatively low-mass halos
contribute to high values of $x$ when $\lambda$ is small, and the
break is smoothed from exponential to power law. Finally, including
the distribution of disk orientations smoothes $h(x)$ further, and
reduces the magnitude of the break in the power law. Figure
\ref{fig-z0p5} also shows that in the standard case with the
full distributions, the shape of $h(x)$ depends only weakly on the
precise value of $\sigma_{\lambda}$ in eq.\ (\ref{eq:spin}).

Also shown in Figure \ref{fig-z0p5} is the effect on $h(x)$ of
changing the parameters $\eta$ and $\zeta$. The effect is simple in a
log-log plot, e.g., multiplying $\eta$ by $10^2=100$ shifts the curve
right by 2 and down by 2, while multiplying $\zeta$ by $100$ shifts
the curve left by 2 and up by 4. More generally, given
$h_{\eta,\zeta}(x)$, changing the parameters to $\eta'$ and $\zeta'$
yields the new function
\be h_{\eta',\zeta'}(x) = \left(\frac{\zeta'}{\zeta}\right)^2 
\frac{\eta} {\eta'}\ h_{\eta,\zeta}\left(\frac{\eta}{\eta'} 
\frac{\zeta'} {\zeta} x \right)\ \label{eq:deg}. \ee As noted above, 
the total cosmic SFR is proportional to $\eta$ and independent of
$\zeta$. The two parameters $\eta$ and $\zeta$ are really effective
fitting parameters. For instance, the parameter $\eta$ includes the
conversion between observed 1500\AA\ flux and underlying SFR, which in
turn includes assumptions about the stellar initial mass
function. Also, any dust extinction reduces the fitted parameter
$\eta$ compared to the actual star formation efficiency, and
extinction could well be significant given the use of rest-frame
ultraviolet wavelengths. The parameter $\zeta$, meanwhile, is exactly
degenerate with the mean spin parameter; varying $\bar{\lambda}$ by a
factor $\alpha$ affects $h(x)$ in the same way as would varying
$\zeta$ by the factor $\alpha^2$. This latter degeneracy can in
principle be lifted by combining measurements of $h(x)$ with other
quantities such as the galaxy luminosity function.

\begin{figure} 
\plotone{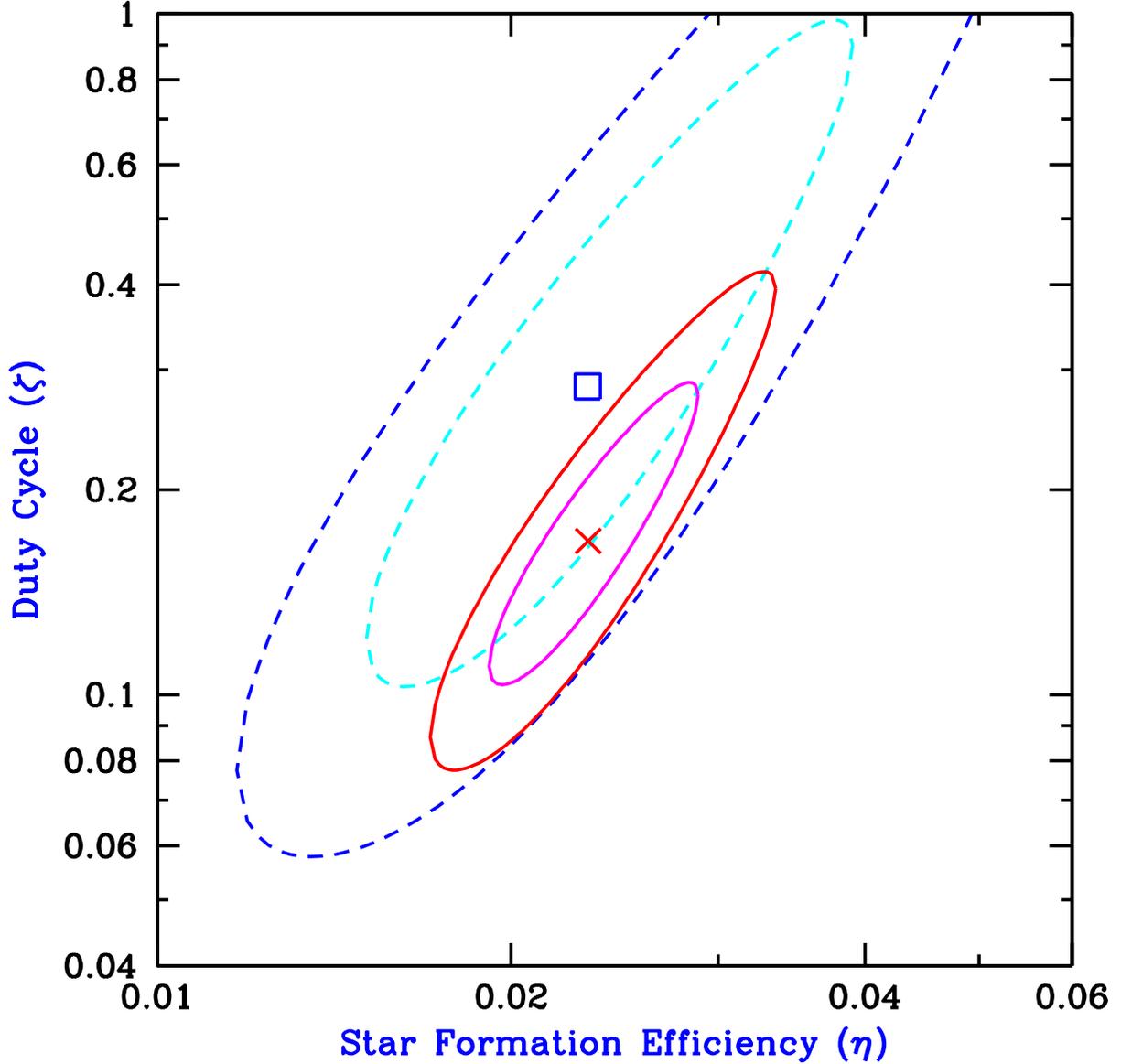} 
\caption{Two-parameter confidence regions, derived from the fitting of
$h(x)$. Shown are constant $\chi^2$ curves in the $(\eta,\zeta)$
space. The $\times$ marks the best-fit point in the fit to the entire
data set, and it is surrounded by solid curves which indicate $68\%$
and $95\%$ confidence limits, respectively. Similarly, the open square
marks the best-fit point in the separate fit to the $z=0.5-1$ bin, and
it is surrounded by dashed curves which indicate $68\%$ and $95\%$
confidence limits, respectively, for this case. Note that we cut off
the figure at the top due to the constraint $\zeta \le 1$.}
\label{fig-2par}
\end{figure}

Eq.\ (\ref{eq:deg}) implies that instead of parameters $\eta$ and
$\zeta$, we can choose two linear combinations (in a log-log plot)
which correspond separately to shifting the $h(x)$ curve right or
left, and shifting up or down. Thus, if $h(x)$ were an exact power
law, there would be an exact degeneracy between the fitted values of
$\eta$ and $\zeta$. Breaking this degeneracy requires a clear feature;
however, the predicted shape of $h(x)$ is an approximate power law
with only a smooth, gradual break. While the power law breaks more
rapidly at high redshift, the $h(x)$ curve is observed over only a
very limited range of $x$ values at high redshift, and thus the
near-degeneracy remains. This degeneracy is clear in Figure
\ref{fig-2par}, which shows two-parameter confidence limits for $\eta$
and $\zeta$. Indeed, the contours are elongated ellipses, and this
limits our ability to separately measure each parameter. Note that the
confidence regions for the fit to the entire data, and the separate
fit to the $z=0.5-1$ bin (the bin with the largest number of data
points), are consistent with each other, with the parameters
constrained much more strongly in the single overall fit.

\begin{figure} 
\plotone{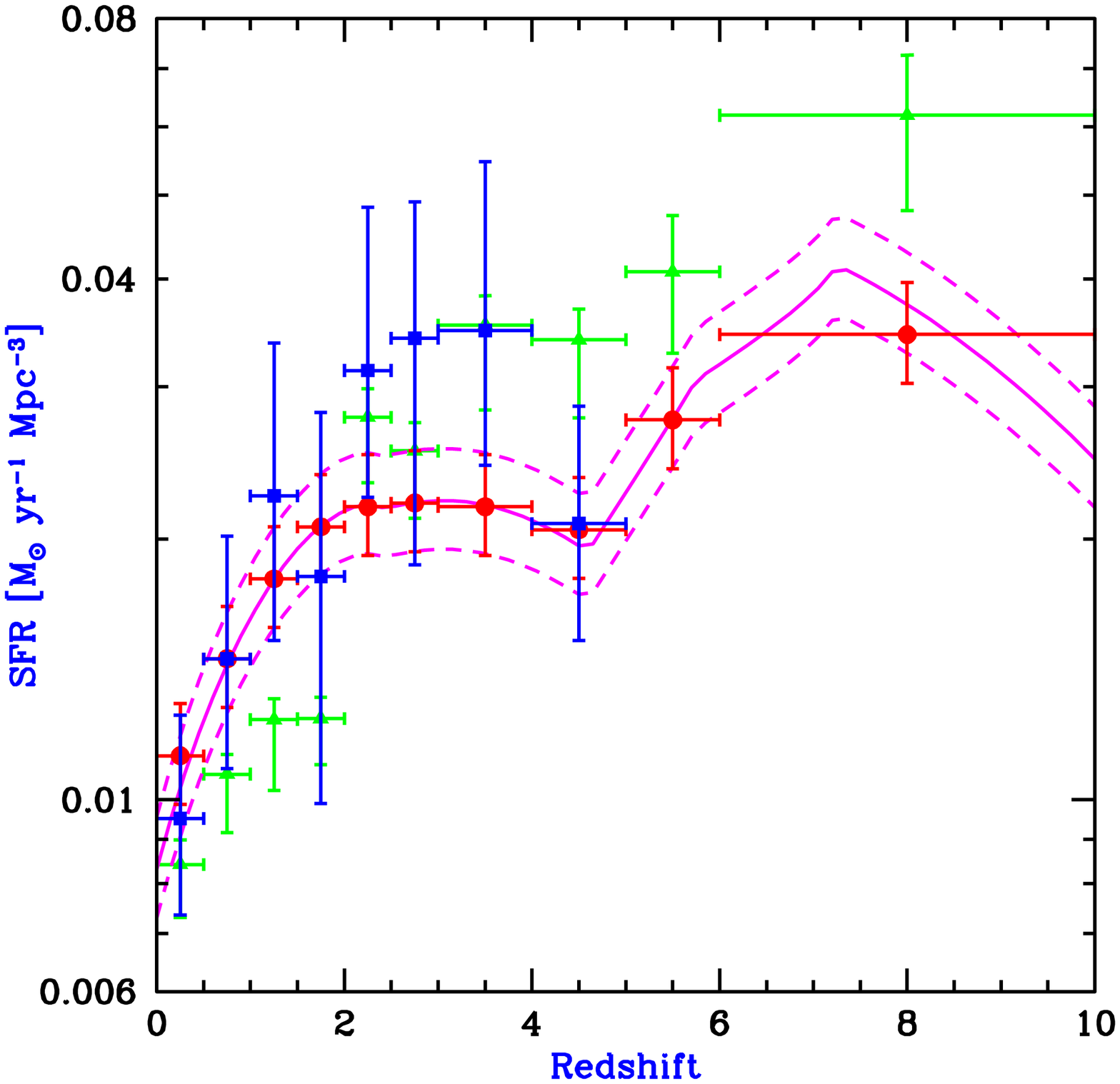} 
\caption{Cosmic SFR versus redshift, derived from the fitting of
$h(x)$. The solid curve shows our best-fit theoretical model, with
dashed curves showing the $1-\sigma$ range. Circles also show $h(x)$
from the theoretical model, but binned in redshift just like the
data. Squares show the theoretical model binned in redshift, but these
are results of independent fits to the $h(x)$ curves within each
redshift bin (The two highest-redshift bins are not shown, since they
contain too few data points to meaningfully constrain our two
parameters). Triangles show the points from Figure 4 of \citet{l02}
[specifically, their green circles, converted into our units and
cosmology].}
\label{fig-SFR}
\end{figure}

Figure \ref{fig-SFR} shows the cosmic SFR versus redshift as derived
from the fitting of $h(x)$. The results of fitting the theoretical
model independently to each redshift bin are consistent, to about 1
$\sigma$, with the result of the simultaneous, overall fit to the
data. Thus, there is no significant indication of an evolution in the
parameters $\eta$ and $\zeta$. Our error bars on the independent fits
in each bin are rather large, caused by the near-degeneracy between
the parameters, as illustrated in Figure \ref{fig-2par}. Note that if
we integrate over the cosmic star formation history, then the model
with $\eta=2.3\%$ implies that the total density parameter in stars
today is $\Omega_* \sim 0.15\%$. This is somewhat lower than the
observed $z=0$ range of $0.19\% < \Omega_* < 0.57\%$ (central value
$0.35\%$; \citealt{fuk98}), and suggests that extinction may be hiding
a substantial fraction of the SFR (although the stellar initial mass
function introduces additional uncertainties into this comparison).

Figure \ref{fig-SFR} also shows one of the fits given by \citet{l02}
to their data. Their results are generally consistent with our
independent fits in each bin, although we do not find evidence for a
large jump in the cosmic SFR at $z=2$. Their error bars are smaller
than ours, but this is a result of their arbitrary decision to allow
only a single parameter to vary in the fit within each redshift
bin. Also, \citet{l02} give two other possible SFR histories from fits
to the data; those fits are much higher than the one shown in Figure
\ref{fig-SFR}, by factors at $z=4$ of roughly 3 and 10, respectively, 
and those extrapolations from the data disagree strongly with our
results, which are based on a physically-derived shape of $h(x)$. The
shape of the cosmic SFR that we find, i.e., a sharp rise from $z=0$ to
$z \sim 1$ followed by a flat segment out to $z \sim 4$, agrees with
estimates based on extrapolations of the galaxy luminosity function
\citep[e.g.,][and other references given in \S 1]{s99}. 

\section{Summary}

We have predicted the distribution of SFR intensity based on models of
hierarchical galaxy formation. We have found that these models provide
a natural explanation for the observed trend of the increase in the
typical SFR intensity with redshift, an increase which occurs despite
the decrease in the typical mass of galaxies. The observed data of
\citet{l02} are consistent with a constant efficiency of turning gas
into stars (best-fit $\eta=2.3\%$) and a constant duty cycle (best-fit
$\zeta=17\%$). Thus, the data are consistent with the standard picture
of the cosmic SFR rising from $z=0$ to $z=2$ and not increasing much
further at $z>2$. 

However, the $h(x)$ data are limited as a probe of star formation
efficiency, since the broad spin-parameter distribution, along with
the distribution of disk orientations, smoothes $h(x)$ into an
approximate power law with only a gradual break. This break is
difficult to measure, since only a limited range of $x$ values can be
detected, especially at high redshift. Thus, two-parameter $\chi^2$
contours show a near-degeneracy between the two fitted parameters, and
much more data would be required in order to fit even more detailed
theoretical models of $h(x)$.

Luminosity functions of galaxies can be obtained for relatively large
populations, using ground-based observations; also, if the total
luminosity of each galaxy can be reliably measured, then this
integrated quantity may allow a more robust comparison with
theoretical models.  However, data in the form of $h(x)$ do have a
number of advantages. The effects of surface-brightness dimming and of
the size distribution of galaxies can both be directly incorporated in
the analysis, unlike the case of galaxy luminosity functions where
both effects enter the selection function in ways that are difficult
to model. In addition, other corrections may also be simpler; e.g.,
extinction may well depend directly on $x$, and not on the total
luminosity of a galaxy, though high resolution images in many
wavebands would be needed in order to determine the level of
extinction separately in every pixel. Indeed, the upcoming Next
Generation Space Telescope should produce a great leap forward for
this type of data. It should probe a much wider range of $x$ values,
with large numbers of galaxies detected over redshifts up to 10 and
beyond. It should also provide high angular resolution over a wide
range of optical and infrared wavelengths, which will likely allow it
to overcome the limitations of current data.

\acknowledgments
This research was supported by an Alon Fellowship at Tel Aviv
University.

\end{document}